\documentclass[reprint, twocolumn,amsmath,amssymb,longbibliography,10pt,aps,pra, superscriptaddress]{revtex4-1}
\usepackage{graphicx}
\usepackage{amsmath,latexsym,tabularx}
\usepackage{xcolor}
\usepackage[%
  colorlinks=true,
  urlcolor=blue,
  linkcolor=blue,
  citecolor=blue
]{hyperref}
\usepackage{booktabs}
\newcommand{\eqnRef}[1]{\mbox{Eq.\,(\ref{#1})}}

\newcommand{\figRef}[1]{\mbox{Fig.\,\ref{#1}}}

\newcommand{\affilLL}[0]{Lincoln Laboratory, Massachusetts Institute of Technology, Lexington, Massachusetts 02421, USA}
\newcommand{\affilMIT}{Massachusetts Institute of Technology, Cambridge, Massachusetts 02139, USA}
\newcommand{\unit}[1]{\, \mathrm{#1}}
\newcommand{\nbardot}[0]{\dot{\bar{n}}}

\begin{document}

\title{Method for Determination of Technical Noise Contributions to Ion Motional Heating}

\author{J. A. Sedlacek}
\affiliation{\affilLL}

\author{J. Stuart}
\affiliation{\affilLL}
\affiliation{\affilMIT}

\author{W. Loh}
\affiliation{\affilLL}

\author{R. McConnell}
\affiliation{\affilLL}

\author{C. D. Bruzewicz}
\affiliation{\affilLL}

\author{J. M. Sage}
\affiliation{\affilLL}

\author{J. Chiaverini}
\affiliation{\affilLL}

\date{\today}

\begin{abstract}
Microfabricated Paul ion traps show tremendous promise for large-scale quantum information processing.  However, motional heating of ions can have a detrimental effect on the fidelity of quantum logic operations in miniaturized, scalable designs. In many experiments, contributions to ion heating due to technical voltage noise present on the static (DC) and radio frequency (RF) electrodes can be overlooked.  We present a reliable method for determining the extent to which  motional heating is dominated by residual voltage noise on the DC or RF electrodes.  Also, we demonstrate that stray DC electric fields can shift the ion position such that technical noise on the RF electrode can significantly contribute to the motional heating rate.  After minimizing the pseudopotential gradient experienced by the ion induced by stray DC electric fields, the motional heating due to RF technical noise can be significantly reduced.
\end{abstract}

\maketitle

\section{Introduction}

Multi-qubit trapped-ion quantum gates rely on coupling of the internal and motional states of the ions. The fidelity of these gates, however, is often limited by decoherence of the motional states due to heating of the ions by electric field noise which can arise due to a number of mechanisms.  One such mechanism is technical voltage noise, which we here define as noise coming from voltage sources, amplifiers, or other electrical components directly connected to the trap electrodes. In some cases, technical noise can be a non-negligible or even primary source of observed ion heating.  In addition to degrading quantum-gate fidelity, this can confound experiments aimed at studying other, more exotic sources of noise~\cite{Brownnutt2015, Turchette2000}. There is thus a need for a method to systematically determine to what extent motional heating is caused by residual technical noise present in experiments.

Here, using a surface-electrode linear Paul trap, we demonstrate a method to determine if experimental motional heating rates are limited by technical noise on static (DC) and radio-frequency (RF) trap electrodes.  The method we employ is based on injecting a known amount of voltage noise onto DC or RF electrodes and measuring the effect on the heating rate.  Noise values over a large range can be investigated with minimal changes to the setup and without changing in-vacuum components, ensuring repeatability of measurements. Additionally, by injecting sufficient noise we can distinctly separate the heating from non-technical sources (often termed ``anomalous'' in cases where the mechanism is unknown) from the heating due to technical noise.  Using this technique, we calculate the contribution to the total heating rate from the amount of measured technical noise present in our system.

Similar work includes Ref.~\cite{Blakestad2009}, which concentrated on measuring the high ion heating rates that arise from a large pseudopotential gradient in the vicinity of a segmented-electrode trap junction.  In the work presented here, however, we demonstrate a technique to investigate heating in a linear Paul trap where pseudopotential gradients are substantially smaller, yet can still cause significant ion heating when other mechanisms are mitigated.  In Ref.~\cite{Schindler2015}, a method using injected noise was employed to qualitatively determine if technical noise was the dominant heating mechanism.  This method did not, however, provide a means to measure the fraction of total ion heating that arose from this technical noise.  In contrast, here we detail a quantitative technique that allows for measurement of this fraction, and thus precisely determines the contribution of technical noise to ion heating rates.  We believe that this methodology can be used broadly to help eliminate technical noise as a limiting source of ion heating.

\section{Theory}

In a linear, RF Paul trap, a single ion can be confined in a three-dimensional harmonic potential with secular frequencies $\omega_i$ $(i \in [x,y,z])$ using a combination of DC and an RF voltages. The RF voltage, oscillating at frequency $\Omega/\omega_i \gg1$ ($\Omega/\omega_i \sim 50$ in our case), generates a pseudopotential that provides trapping in the two radial directions orthogonal to the long trap axis. The DC voltages create trapping fields along the axial direction  \cite{Leibfried2003,theBible}. Ideally, the ion can be trapped at the minimum of the pseudopotential where the pseudopotential gradient is zero; this location is often referred to as the RF null.  However, in the presence of stray electric fields, the ion will be shifted away from the pseudopotential minimum.  In general, this results in the ion being located in a region where the pseudopotential gradient is nonzero in both the radial and axial directions, with the axial gradient arising due to, e.g., the asymmetry and finiteness of the trap.

The motional heating rate $\dot{\bar{n}}_{i}$ experienced by a trapped ion along direction $\vec{\textit{\i}}$ is sensitive to electric field noise at its secular frequency, as well as to noise on the pseudopotential at frequencies $\Omega \pm \omega_i$.  This electric field noise is typically coupled to the ion through voltage noise present on the trap electrodes.  These technical noise contributions to the heating rate can be expressed as \cite{Brownnutt2015, Blakestad2009},
\begin{equation}
\begin{split}
\dot{\bar{n}}_{i,\mathrm{tech}} & = \frac{q^2}{4 m \hbar \omega_i} \bigg( \sum_j  \frac{S_{V_{j}}(\omega_i)}{D_{i,j}^2} \\
& + \frac{q^2}{4 m^2 \Omega^4} \left[ \frac{\partial}{\partial i} E_0^2 \right ]^2 \frac{S_{V_\mathrm{RF}}(\Omega \pm \omega_i)}{V_0^2} \bigg)
\end{split}
\label{eq:hr1}
\end{equation}
where $q$ is the charge of the ion, $m$ is the mass of the ion, $S_{V_j}(\omega_i)$ is the voltage noise spectral density on the $j$th electrode, $E_0^2$ is the square of the magnitude of the electric field $\overrightarrow{E_0}(x,y,z)$ at frequency $\Omega$, produced by the RF electrode at the equilibrium position of the ion, $S_{V_\mathrm{RF}}(\Omega \pm \omega_i)$ is the voltage noise spectral density on the RF electrode, and $V_0$ is the peak driving voltage applied to the RF electrode. The characteristic distance $D_{i,j}$ is calculated by dividing the voltage applied to the $j$th electrode $V_j$ by the resulting electric field $E_{i}$ along the principal axis $i$ at the equilibrium position of the ion \cite{Brownnutt2015,Leibrandt2007},

\begin{equation}
D_{i,j} = \frac{V_j}{E_{i}}.
\label{eq:Deff}
\end{equation}  

\begin{figure}[h t !]
\includegraphics[width = \columnwidth]{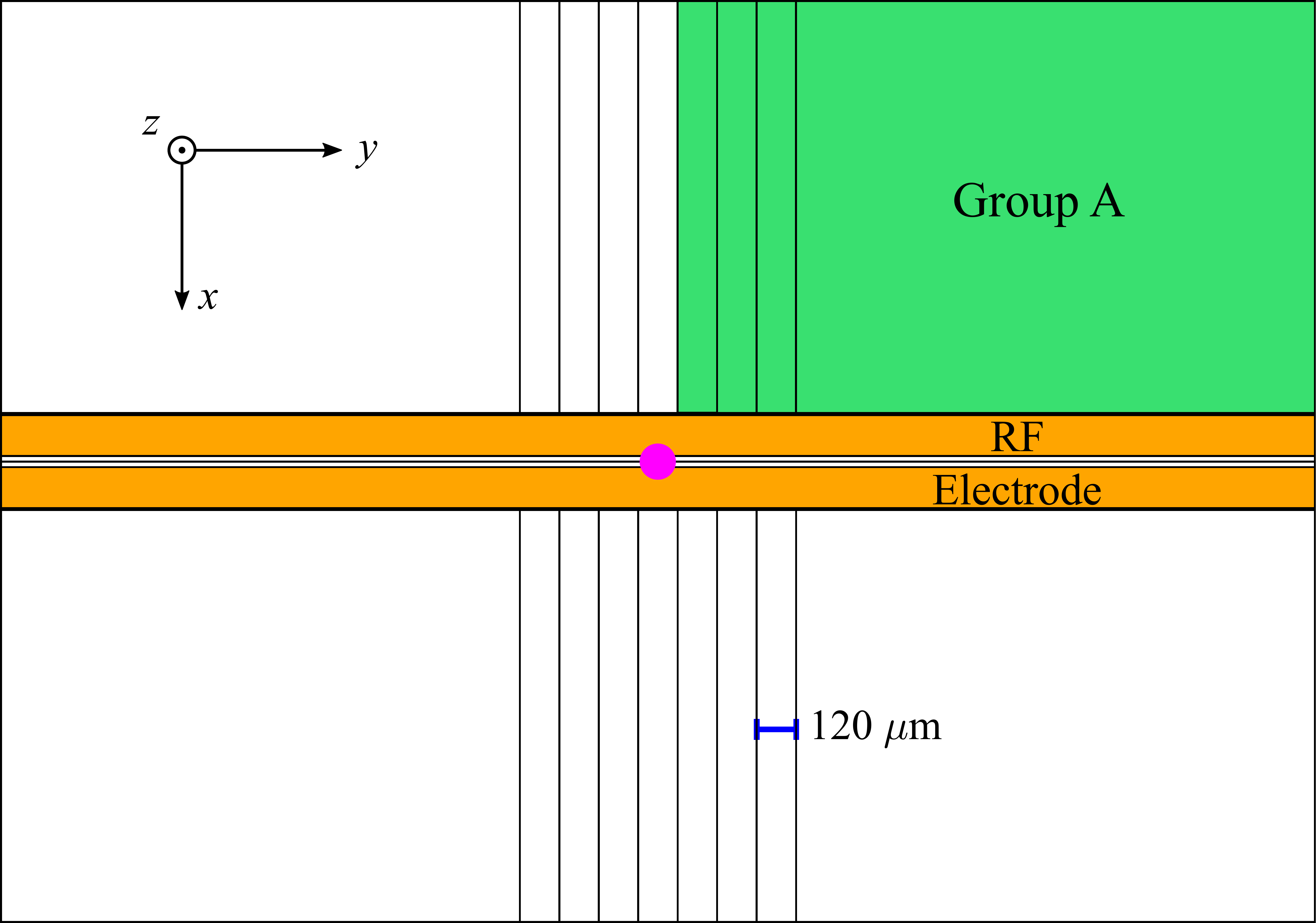}
\caption{Schematic of the electrodes in the central region of the surface-electrode trap.  Each group of shaded electrodes is used independently for noise injection.  A $^{88}\mathrm{Sr}^+$ ion is trapped $\sim 50 \, \mu \mathrm{m}$ from the electrode surface.  The pitch of the segmented central electrodes is $120 \unit{\mu m}$.  The solid black lines denote gaps in the metal of width $\sim 5 \unit{\mu m}$. The position of the ion is approximated by the purple dot, and is not to scale.}
\label{fig:electrodes}
\end{figure}

\noindent The field component $E_{i}$ is calculated using an electrostatic boundary element simulation using the trap geometry shown in \figRef{fig:electrodes}, after applying $V_{j}$ to electrode $j$ and grounding all other electrodes.  Alternatively, as will be shown below, $D_{i,j}$ can be determined experimentally via noise injection experiments.

The first term in \eqnRef{eq:hr1} illustrates the dependence of the heating rate on noise at the secular frequency $\omega_i$.  The voltage noise on each electrode causes electric field fluctuations at the position of the ion with the noise spectral density of the electric field, $S_{E_j}(\omega_i) = S_{V_j}(\omega_i)/D_{i,j}^2$.  Anomalous heating is typically attributed to this mechanism. 

The second term in \eqnRef{eq:hr1} contains the expression $\frac{\partial}{\partial i} E_0^2$, which is proportional to the gradient of the pseudopotential.  Technical noise near the RF frequency can heat the motion of a particular mode if the gradient of the pseudopotential at the ion location has a nonzero component along that mode's primary axis.  In a typical linear trap, this expression becomes important when the ion is displaced away from the RF null.

\begin{figure}[h t !]
\includegraphics[width = \columnwidth]{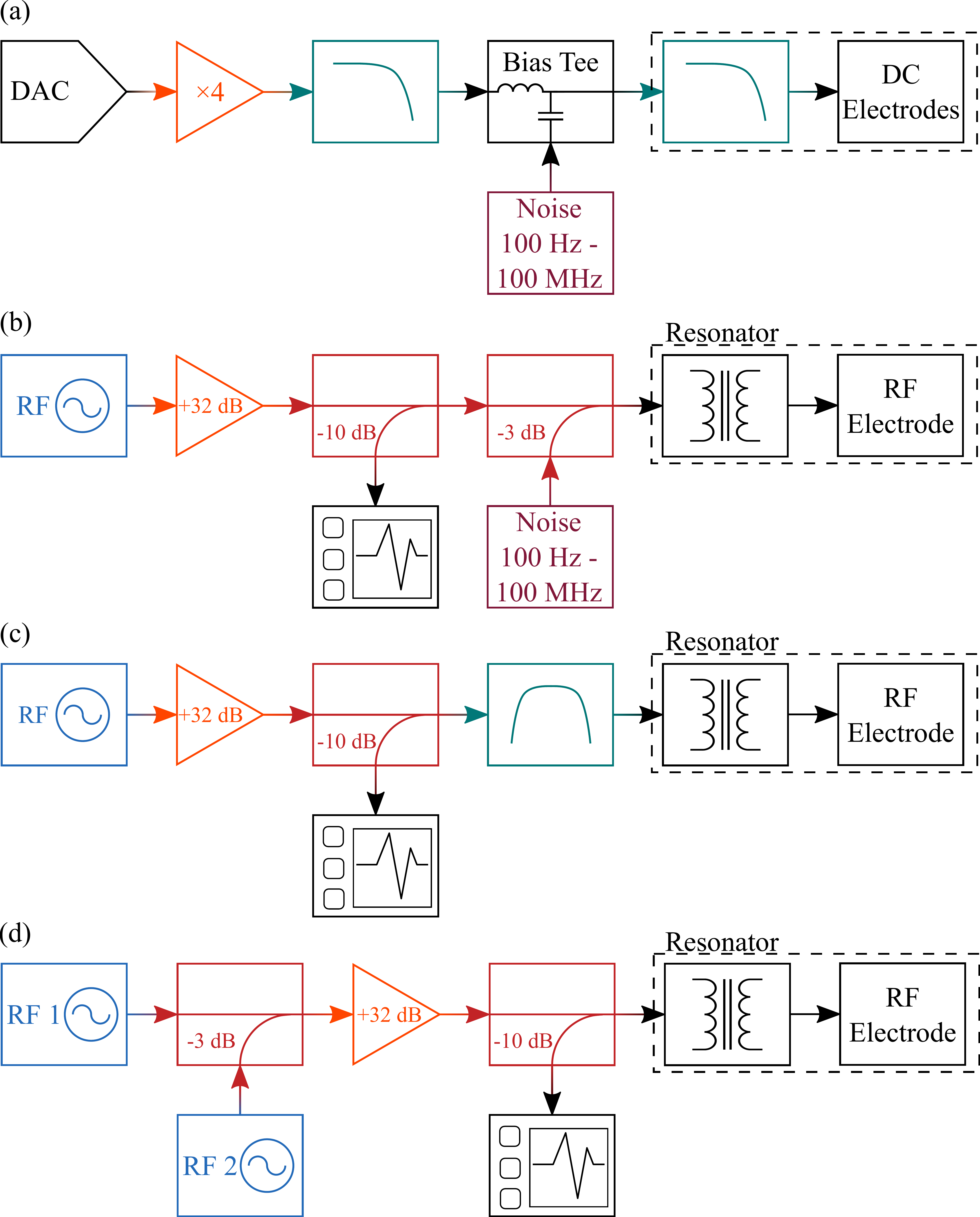}
\caption{Diagrams for the RF and DC voltage circuits, illustrating how noise is injected or filtered.  Items inside the dashed blocks indicate electronics that are inside the vacuum system.  (a) Circuit for applying DC voltages to the electrodes, while injecting noise.  The DC voltages are produced by digital to analog converters (DACs), (NI PXI-6723).  The voltages are amplified by a factor of 4 to increase the output range and are filtered with low-pass filters (Kiwa Electronics) that provide $> 95 \, \mathrm{dB}$ of filtering at $\omega_{y}=2\pi\times1.29$~MHz.  Noise is added to the DC voltages using a bias tee (Minicircuits ZFBT-6GW).  Inside vacuum near the trap electrodes are RC low-pass filters with a cutoff frequency of $\sim 2.5 \unit{kHz}$.  (b) Diagram illustrating the addition of noise to the RF voltage.  The RF source (SRS SG 384) is amplified by +32 dB (Minicircuits ZHL-1-2W).  A directional coupler is used to monitor the signal that is reflected from the resonator.  The noise is combined with the RF trapping voltage using a power combiner (3 dB coupler).  Inside the vacuum chamber a helical resonator is used to impedance match the circuit with the RF electrode.  (c) Schematic for the filtering of the RF voltage.  The circuit is the same as in (b), except the noise and the power combiner are replaced by two tunable bandpass filters.  The bandpass filters (K\&L Microwave 5BT-30/76-5-N/N) are tuned to let through frequency $\Omega$ and filter out noise at $\Omega \pm \omega_y$ by $20 \unit{dB}$.  (d) Schematic for the setup used to minimize the pseudopotential gradient induced by a stray electric field. RF source 2 is used to excite the ion at $\Omega + \omega_y$.  Additional details are contained in the text.}
\label{fig:noiseInjection}
\end{figure}

\section{Experiment}

The details of the ion trapping apparatus have been described previously \cite{Sage2012, Chiaverini2014,McConnel2015,Bruzewicz2015}.  Single $^{88}\mathrm{Sr}^+$ ions are trapped $50 \unit{\mu m}$ above a surface-electrode trap cooled to $4 \unit{K}$ by a closed-cycle cryocooler. The cryocooler and an ion pump provide an ultra high vacuum environment for ion trapping.  Sr$^{+}$ ions are loaded by photoionizing neutral Sr that is precooled in a 2D magneto-optical trap located in a separate differentially pumped vacuum chamber \cite{bruzewicz2016scalable}.

The trap chip used in these experiments is a segmented linear Paul trap, and the trap electrodes are 2~$\mathrm{\mu m}$ thick aluminum sputtered onto a sapphire substrate.  A schematic of the central region of the electrodes is shown in \figRef{fig:electrodes}.  The RF~electrode, which splits into two parallel traces on the chip (with other electrodes between the two branches), is highlighted in yellow; all other electrodes are DC electrodes.  In this work, we focus on heating of the ion's axial mode ($i = y$).  

Noise injection is carried out in two different cases, either on DC or RF electrodes.  The noise source we use is a Noisecom 6107, which generates a flat noise spectrum from $100 \unit{Hz} - 100 \unit{MHz}$.  A broadband source is advantageous to ensure excitation of the ion at its secular frequency while also allowing for exploration of any other frequencies that may cause heating; the noise spectrum can be tailored by filtering different frequency bands.

The electrical schematic for the noise injection to the DC electrodes is shown in \figRef{fig:noiseInjection}a.  The noise is added to an amplified and filtered DC control voltage generated by a 13-bit digital-to-analog converter (NI PXI-6723),  using a bias tee.  The combined signal is sent to a set of four electrodes that are electrically connected together on chip.  They are denoted as \mbox{Group A} in \figRef{fig:electrodes}.

The electrical schematic for the noise injection onto the RF electrode is shown in \figRef{fig:noiseInjection}b.  The RF source is a SRS SG 384, and is amplified (+32 dB) by a Minicircuits ZHL-1-2W broadband amplifier.  The noise is combined with the RF drive using a power combiner.  A voltage step-up helical resonator is used to impedance match the circuit to the RF electrode. In lieu of injecting noise, bandpass filtering the RF signal centered around $\Omega$ was done as shown in \figRef{fig:noiseInjection}c in some experiments.  The bandpass filters reduce the noise at $\Omega \pm \omega_{y}$ by $20 \unit{dB}$ for $\omega_{y} = 2\pi\times1.29 \unit{MHz}$.     

To investigate the influence of stray electric fields shifting the ion into a position with a higher pseudopotential gradient, a method for minimizing the gradient was required.  To do this a second RF source at $\Omega + \omega_y$ is used to drive the axial motion of the ion as shown in \figRef{fig:noiseInjection}d; note the similarity of the circuits used for gradient minimization(\figRef{fig:noiseInjection}d) and noise injection on the RF electrode(\figRef{fig:noiseInjection}b).  In both methods, the ion is being excited by an external signal through the same mechanism, the second term in \eqnRef{eq:hr1}.  (In the case of white noise injection, the ion's motional state satisfies a thermal distribution, while for gradient minimization, this drive excites the ion into a nonthermal state.)  As a result, minimizing the ion's motional excitation, and thus the gradient from the RF source, will also minimize the excitation of the ion from technical and injected noise.  

The frequency of the second RF source is swept across $\Omega + \omega_y$ while the motional excitation of the ion is observed on an EMCCD camera.  The ion is translated in the $x$, $y$, and $z$ directions by adjusting the DC voltages to minimize the amount of induced excitation of the ion.  After gradient minimization, no excitation is observed even for injected driving power equal in magnitude to the trapping RF tone. We find that the $z$-direction adjustment has the largest effect on the minimization of the excitation.

Note that the above experimental setup and procedure for pseudopotential gradient minimization is similar to finding the RF null through parametric excitation of the radial modes \cite{Ibaraki2011}, however they are not equivalent. The method of parametric excitation relies on modulating the pseudopotential to heat the ion, where the pseudopotential is the main contributor to the total potential \cite{Tanaka2012}. However, in the case of a linear Paul trap, the axial potential is determined predominately by the DC potential, and modulating the RF pseudopotential does not result in parametric excitation along the axial direction. This was verified in our case by measuring the amplitudes of RF drive required to excite equivalent motion at $\Omega + \omega_y$ and $\Omega + 2 \omega_y$. The strongest response for a parametric oscillator is expected for driving at frequency $\Omega + 2 \omega_y$. However, when using the gradient minimization method, $> 30 \unit{dB}$ of increased drive power was needed to excite an ion the same amount at $\Omega + 2 \omega_y$ when compared to $\Omega + \omega_y$.

Heating rates were measured using the method of sideband spectroscopy~\cite{Chiaverini2014, Monroe1995}.  Using sideband cooling, the ion is prepared in the axial motional ground state of the $\vert S_{1/2}, m_J = -1/2 \rangle$ level.  After a variable delay time, the average vibrational occupation $\bar{n}$ is determined by measuring the amplitudes of the red and blue sidebands.  Heating rate measurements were carried out with an axial trapping frequency of $\omega_y=2\pi\times1.29 \unit{MHz}$ using various amounts of injected noise.

By injecting a known amount of noise onto electrodes, measuring the increased heating rate of the ion allows us to measure the transfer function of injected voltage noise on the electrodes to the motional heating rate of the ion. The aim of these experiments is to determine the contribution of residual technical noise to the intrinsic heating rate, which is the heating rate taken under normal experimental conditions without any injected noise. Once this transfer function is known, we can calculate the amount of heating caused by the independently measured residual technical noise in the system.  To isolate the effects of injected noise on heating rates, first an intrinsic heating rate is measured, i.e. without any injected noise. The intrinsic heating rate could be due to a variety of sources, including electromagnetic interference, anomalous heating \cite{Brownnutt2015}, and any residual technical noise in the system.  Afterwards, a series of heating rate measurements are made with various amounts of injected noise on either the RF or DC electrodes.  The level of noise was chosen such that heating due to the injected noise was the dominant heating mechanism.

In the presence of injected noise, the total heating rate of the ion can be written as

\begin{equation}
\nbardot_\mathrm{tot}(\omega_{i}) = \nbardot_\mathrm{bg} + \frac{q^2}{4 m \hslash \omega_i} \sum_j \frac{S_{V_{j}}^{(0)}(\omega_i)+S_{V_{j}}^{(\mathrm{inj})}(\omega_i)}{D_{i,j}^2}
\label{eq:omegaFits}
\end{equation}

\noindent where $\dot{\bar{n}}_{\mathrm{bg}}$ is the background contribution to the heating rate, which is different from the intrinsic heating rate as it does not include the contribution from residual technical noise. The sum is over electrodes $j$ and includes the technical noise contribution from any injected noise at frequency $\omega_i$ on the $j$th electrode $S_{V_{j}}^{(\mathrm{inj})}(\omega_i)$ as well as residual technical noise present on that electrode $S_{V_{j}}^{(0)}(\omega_i)$. Experimental data is fit to \eqnRef{eq:omegaFits} using the experimentally measured $S_{V_{j}}(\omega_i)=S_{V_{j}}^{(0)}(\omega_i)+S_{V_{j}}^{(\mathrm{inj})}(\omega_i)$, with the free parameters $\nbardot_\mathrm{bg}$ and $D_{i,j}^2$ (we determine the characteristic distance here experimentally and compare it with the value extracted from potential simulation).  

Similarly, for the voltage noise at $\Omega \pm \omega_i$ the total heating rate is
\begin{equation}
\begin{split}
\nbardot_\mathrm{tot}(\Omega \pm \omega_{i}) = & \nbardot_\mathrm{bg} + \frac{q^4}{16 m^3 \hslash \Omega^4 \omega_i} \left[ \frac{\partial}{\partial i} E_0^2 \right ]^2 \\
& \times \frac{S_{V_\mathrm{RF}}^{(0)}(\Omega \pm \omega_i) + S_{V_\mathrm{RF}}^{(\mathrm{inj})}(\Omega \pm \omega_i)}{V_0^2},
\end{split}
\label{eq:bigOmegaFits}
\end{equation}
where the injected noise on the RF electrode at frequencies $\Omega \pm \omega_i$, $S_{V_\mathrm{RF}}^{(\mathrm{inj})}(\Omega \pm \omega_i)$, is summed with the residual noise on the RF electrode, $S_{V_\mathrm{RF}}^{(0)}(\Omega \pm \omega_i)$. \eqnRef{eq:bigOmegaFits} is fit to experimental data with the free parameters $\nbardot_\mathrm{bg}$ and $\frac{\partial}{\partial i} E_0^2$.  

By varying the amount of injected noise on the DC electrodes and fitting the measured heating rates $\nbardot_\mathrm{tot}(\omega_{i})$ as a function of the total voltage noise on electrodes, $S_{V_{j}}(\omega_i) = S_{V_{j}}^{(\mathrm{0})}(\omega_i)+S_{V_{j}}^{(\mathrm{inj})}(\omega_i)$, we determine the mapping of $S_{V_{j}}(\omega_i)$ to ion axial heating rate.  After taking the limit as $S_{V_{j}}^{(\mathrm{inj})}(\omega_{i}) \rightarrow 0$, and $\dot{\bar{n}}_{\mathrm{bg}} \rightarrow 0$ in \eqnRef{eq:omegaFits}, and using the measured $S_{V_{j}}^{(\mathrm{0})}(\omega_i)$ in the absence of injected noise, we determine the contribution to the intrinsic heating rate for electrode $j$ due to residual (non-intentional) sources of technical noise. The intrinsic heating rate is defined as $\nbardot_\mathrm{int}(\omega_{i}) = \nbardot_\mathrm{tot}(\omega_{i}, S_{V_{j}}^{(\mathrm{inj})} \rightarrow 0)$. An analogous technique is used for noise on the RF electrode.  This method is in general applicable to both axial and radial modes; in the following sections we present data for motional heating of the axial mode, $i=y$.  

\begin{figure}[t]
\includegraphics[width = \columnwidth]{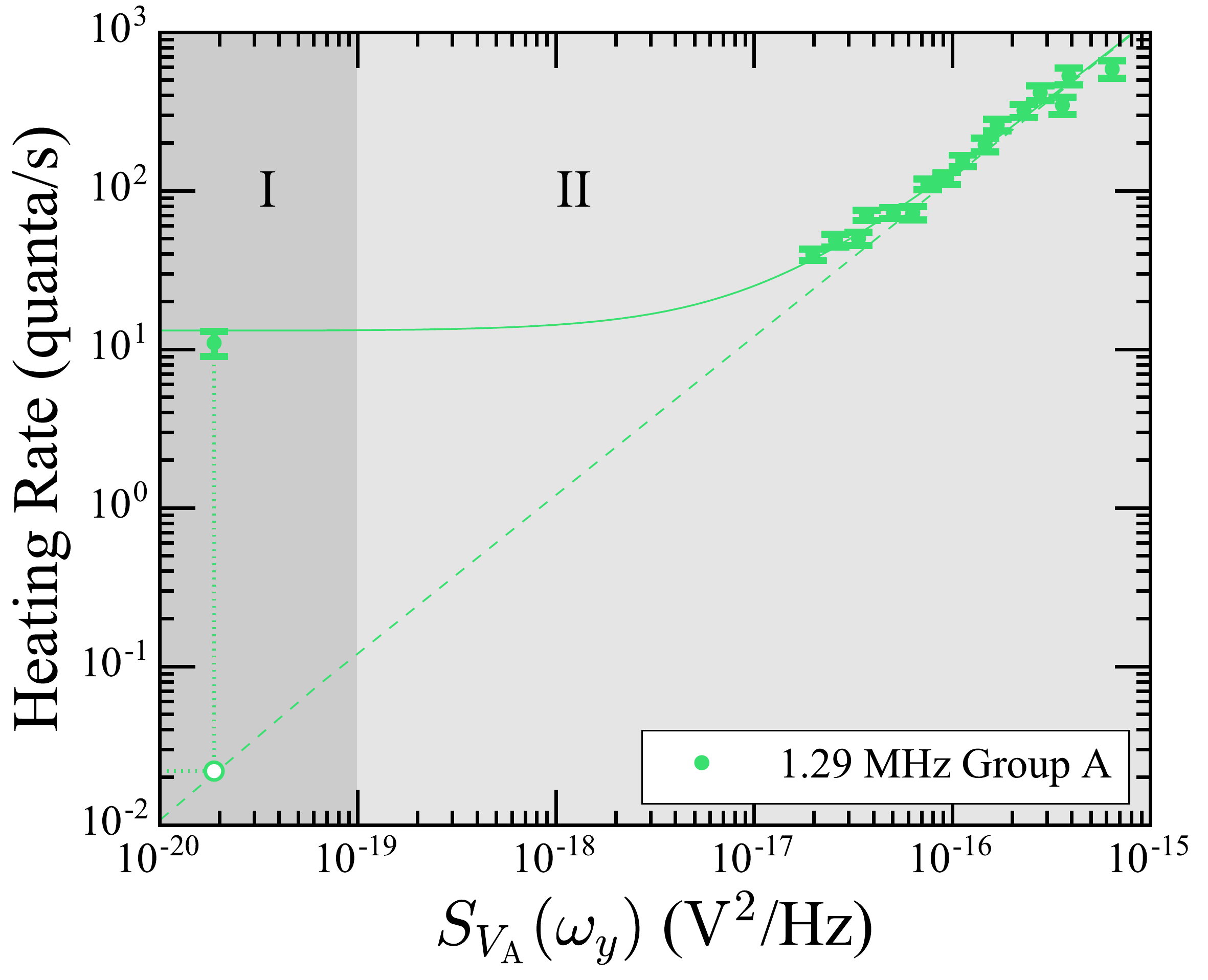}
\caption{Heating rates for various amounts of voltage noise, $S_{V_\text{A}}(\omega_y)$ as measured on the DC electrodes. The solid line is a fit to \eqnRef{eq:omegaFits}. The dashed line is a plot of \eqnRef{eq:omegaFits} with $\nbardot_\mathrm{bg}=0$ showing the contribution of voltage noise on \mbox{Group A} electrodes to the total heating rate.  The data point in region~I is the intrinsic heating rate, and was taken without noise injection.  The data in region II were taken with injected noise.  The x-values are calculated by measuring the voltage noise outside of the chamber and multiplying by the measured response of the in-vacuum filters. To find the heating rate contribution due to residual technical noise, the dashed line is evaluated at the measured level of residual noise $S_{V_\text{A}}(\omega_y)$. This is illustrated by the dotted lines and open circle.}
\label{fig:dcNoiseInjection}
\end{figure}

\section{Results}

\subsection{Noise Injection on DC electrodes}
Noise was injected on a set of four electrodes held at the same voltage, shown as \mbox{Group A} in \figRef{fig:electrodes}. \mbox{Group A} electrodes were chosen because they have the largest electric field component along the axial direction in comparison to other electrodes.  Hence, noise on these electrodes had the largest contribution to the heating rate.  

To calculate the voltage noise on the DC trap electrodes, the noise is first measured outside the chamber using a spectrum analyzer.  This value is then multiplied by the response, measured separately, of the on-board RC filter at the trap frequencies to determine the noise $S_{V_\text{A}}(\omega_y)$ on the electrodes.  The transfer function of the RC filter is determined  by measuring the amount of attenuation of an AC signal through the filter on a return line outside of the chamber.

\begin{figure}[!tbp]
\includegraphics[width = \columnwidth]{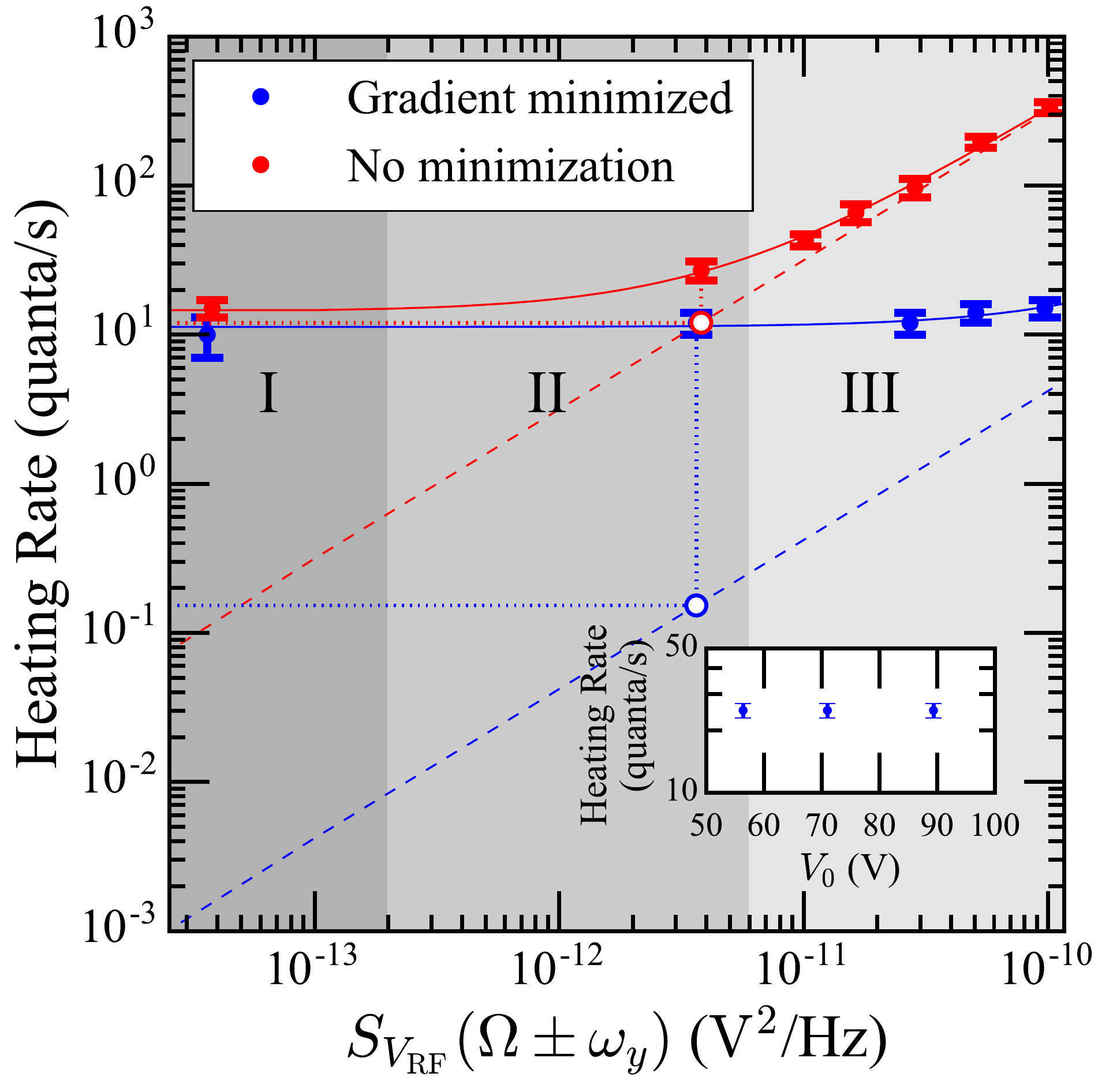}
\caption{Measured heating rate for various amounts of RF noise with a trap frequency of $1.29 \unit{MHz}$.  $S_{V_\mathrm{RF}}(\Omega \pm \omega_y)$ is the voltage noise on the RF electrode derived from measurements (as explained in the text).  The red and blue solid lines are fits to \eqnRef{eq:bigOmegaFits}.  The dashed lines show the contribution of the of the voltage noise at $\Omega \pm \omega_y$ to the total heating rate.  Regions I, II, and III correspond to cases of extra bandpass filters, residual noise only, and injected noise, respectively.  Open circles and dotted lines denote the expected contributions to the intrinsic heating rate from the measured residual technical noise. The inset in (b) shows the measured heating rate for different values of trap voltage amplitude $V_0$ on the RF electrode. }
\label{fig:rfNoiseInjection}
\end{figure}

Heating rates from the noise injection experiments are shown in \figRef{fig:dcNoiseInjection}.  The data point in region~I was taken with no injected noise and corresponds to $\nbardot_\mathrm{int}(\omega_{y})$, while the data in region II was taken with injected noise.  The solid line is a fit to \eqnRef{eq:omegaFits} using the measured values of $S_{V_\text{A}}(\omega_y)$ with the free parameters of $\dot{\bar{n}}_\mathrm{bg}$ and $D_{y,\mathrm{A}}$. The dashed line is a plot of \eqnRef{eq:omegaFits}, with $\nbardot_\mathrm{bg} = 0$. The line indicates the contribution of voltage noise on Group~A to the total heating rate. Using the value of measured residual noise, the contribution of Group~A to $\nbardot_\mathrm{tot}(\omega)$ is $0.023 \unit{quanta/s}$. Since there are four groups on the trap, for uncorrelated noise, the contribution to the intrinsic heating rate is $0.092 \unit{quanta/s}$, less than \textbf{$1 \%$} of the intrinsic heating rate.  For noise correlated and in phase on the four quadrants, we expect no heating by symmetry.  In the very unlikely case that noise was anti-correlated in phase for quadrants with positions of opposite location in~$y$ with respect to the ion location, we would expect a maximum contribution of $0.37 \unit{quanta/s}$, less than \textbf{$4 \%$} of the intrinsic heating rate.  Though the contribution from intrinsic technical noise is small in our case, in situations where it is much larger, the extent of noise correlation between electrodes should be considered.

We have also calculated the expected Johnson noise on the DC electrodes, originating from the in-vacuum RC low-pass filters, and the resulting motional heating. Following the calculations outlined in Ref. \cite{Brownnutt2015}, the expected Johnson noise is $S_{V_{A}}(\omega_y) \sim 1 \times 10^{-22} \unit{V^2/Hz}$, leading to a total expected contribution to the heating rate of $\sim 10^{-4} \unit{quanta/s}$.

Next we compare the characteristic distance determined from ion heating rate measurements with that calculated separately.  Since the electrodes in Group~A were connected together and shared the same potential, one characteristic distance was calculated by treating the four separate electrodes as a single electrode.  From the fit to \eqnRef{eq:omegaFits}, $D_{y,\mathrm{A}} = 6.5(1) \unit{mm}$. From trap simulations using \eqnRef{eq:Deff}, $D_{y,\mathrm{A}} = 6.4 \substack{+0.2 \\ -0.6}\unit{mm}$ where the uncertainty is due to the uncertainty in ion position.  The agreement between the experimental measurement and the theoretical expectation for $D_{y,\mathrm{A}}$ demonstrates that detailed noise injection measurements on DC electrodes may not be necessary to determine if technical noise on DC electrodes is a major contribution to the total heating rate.  Noise measurements outside of vacuum, known transfer functions of in-vacuum components, and calculation of $D_{i,j}$ are sufficient to determine the amount of motional heating expected by this type of technical noise. On the other hand, noise injection measurements will be useful if $D_{i,j}$ is not known or cannot be calculated accurately, or as a method to verify that the value $D_{i,j}$ has been determined correctly.  

Using a low pass filter with a cutoff frequency of $1.9 \unit{MHz}$ to filter the injected noise, no change in the heating rate was measured, as compared to the heating rate while injecting broadband noise without any low pass filter.  This indicates that the increase in heating rate is due solely to the noise at the axial frequency $\omega_y$ and not at the radial mode frequencies, which are greater than $2 \unit{MHz}$, or to noise at $\Omega \pm \omega_y$. Since the main contributor to the increase in heating rate is noise at the axial frequency $\omega_y$, the use of \eqnRef{eq:omegaFits} is justified.  In fact, we verify the use of \eqnRef{eq:omegaFits} in general.

\subsection{Noise Injection on RF Electrode}
We measured heating rates for a range of noise amplitudes on the RF electrode. The noise on the RF electrode \mbox{$S_{V_\mathrm{RF}}(\Omega \pm \omega_y) = S_{V_\mathrm{RF}}^{(0)}(\Omega \pm \omega_y) + S_{V_\mathrm{RF}}^{(\mathrm{inj})}(\Omega \pm \omega_y)$} is calculated in a similar manner to the noise on the DC electrodes.  The power spectral density of the noise $S_P(\Omega+\omega_y)$ is measured using a spectrum analyzer on the input to the electrical vacuum feedthrough that feeds the RF signal to the helical resonator (held at $\sim50\unit{K}$) located inside the vacuum chamber. $S_P(\Omega+\omega_y)$ is then multiplied by the response of the resonator to calculate the total voltage noise on the RF electrode,
\begin{equation}
S_{V_\mathrm{RF}}(\Omega + \omega_y) = \frac{Q L \Omega}{1+4 Q^2 \frac{\omega_y^2}{\Omega^2}} S_P(\Omega+\omega_y),
\label{eq:rfNoys}
\end{equation}
where $Q=170$ is the measured quality factor of the resonator, and $L = 500 \unit{nH}$ is the inductance of the resonator.  The Lorentzian accounts for attenuation of transmitted noise power away from resonance $\Omega$.  Experimentally we observe $S_P(\Omega+\omega_y) = S_P(\Omega-\omega_y)$, and \eqnRef{eq:rfNoys} is multiplied by a factor of two to calculate the total voltage noise $S_{V_\mathrm{RF}}(\Omega \pm \omega_y)=S_{V_\mathrm{RF}}(\Omega + \omega_y) +S_{V_\mathrm{RF}}(\Omega - \omega_y) $ on the RF electrode.

The results of noise injection on the RF electrode for a trap frequency of $1.29 \unit{MHz}$ are shown in \figRef{fig:rfNoiseInjection}. Heating rates were measured under three conditions: first, bandpass filters were used to filter out noise on the RF source at $\Omega \pm \omega_y$, as displayed in Region I and as detailed in \figRef{fig:noiseInjection}c; second, nominal conditions were used with no filters and no injection of noise (intrinsic heating rates with residual noise), as displayed in Region II; third,  noise was injected, as displayed in Region III and as detailed in \figRef{fig:noiseInjection}b.

The solid red and blue lines are fits to \eqnRef{eq:bigOmegaFits} with the free parameters $\dot{\bar{n}}_{\mathrm{bg}}$ and $\frac{\partial}{\partial y} E_0^2$.  The dashed red and blue lines are plots of \eqnRef{eq:bigOmegaFits} with $\nbardot_\mathrm{bg} = 0$, showing the contribution of $S_{V_\mathrm{RF}}(\Omega \pm \omega_y)$ to the total heating rate.  Data was taken with (blue) and without (red) gradient minimization.  Such compensation reduces the sensitivity of the ion to injected noise.

For the case of no gradient minimization (shown in red) the fit to \eqnRef{eq:bigOmegaFits} yields $\frac{\partial}{\partial y} E_0^2 = 2.1(5) \times 10^{12} \unit{V^2/m^3}$, and with gradient minimization shown in blue, $\frac{\partial}{\partial y} E_0^2 = 2(2) \times 10^{11} \unit{V^2/m^3}$.  Since $\nbardot \propto ( \frac{\partial}{\partial y} E_0^2)^2$, a factor of 100 less noise is coupled to the ion through the pseudopotential after minimization.  In addition, the fitted gradient is consistent with zero after gradient minimization was performed. 
The largest gradient measured here, in the uncompensated case, is two orders of magnitude smaller than the largest gradient measured in Ref.~\cite{Blakestad2009}, $\frac{\partial}{\partial y} E_0^2\sim\nobreak 2\times 10^{14} \unit{V^2/m^3}$.  This difference is expected due to the differing trap geometries, but illustrates that even in the case of a linear trap, pseudopotential gradients cannot always be ignored.       

Trap-potential simulations show many points in the vicinity of the RF null where the pseudopotential gradient along the axial direction is zero.  Due to our method of compensating for stray fields, which results in localization of the ion to one of these minima of the axial pseudopotential gradient and not necessarily to the position of the RF null, we are unable to precisely determine the ion position relative to the electrodes.  This prevents us from comparing the data with simulated pseudopotential gradients; moreover the gradient may be difficult to calculate accurately in general, for example when it is very small in comparison with field gradients that arise from wirebonds or fabrication imperfections.  In light of this, our method, which does not require accurate {\it a priori} knowledge of the pseudopotential gradient, is viable to experimentally determine sensitivities to noisy RF signals.

Without gradient minimization, the noise on the typical RF source we use (SRS SG 384) is enough to contribute significantly to the intrinsic motional heating rate, as shown in Region II. The power spectral density of the noise is measured to be \mbox{$S_P(\Omega+\omega_y)= 8 \times 10^{-15} \unit{W/Hz}$} outside the chamber before the helical resonator. Technical noise contributes $12(1) \unit{quanta/s}$ to the intrinsic heating rate of $27(4) \unit{quanta/s}$. After compensating the stray electric field, the contributions of technical noise are reduced to negligible levels.  In this case, technical noise contributes $0.15(4) \unit{quanta/s}$ to an intrinsic heating rate of $12(2) \unit{quanta/s}$. These are the contributions to the heating rates under our usual operating conditions without any bandpass filters on the RF circuit. The contributions to the intrinsic heating rate will be even smaller with the addition of the bandpass filters.

In addition to using bandpass filters, we have also found two other methods for reducing motional heating rates caused by technical noise.  Using a different RF generator with $> 20 \unit{dB}$ less noise (HP 8640B) results in a heating rate consistent with heating rates taken while using the SG 384 and the bandpass filters.  Additionally, we have found that the noise from the SG 384 is independent of output RF power.  Increasing the output power of the RF generator by 10 dB and putting a 10 dB attenuator on the output of the generator also led to a reduction in heating rate equivalent to using the bandpass filters.  

The bandpass filters have also been used to select different frequency components of the injected noise on the RF electrode. It was found that filtering noise outside of $\Omega \pm \omega_y$ did not lead to an observable reduction of the heating rate as compared to using broadband noise. This rules out the possibility of noise at the secular frequencies $\omega_i$ causing the increased heating.  Additionally, filtering the injected noise to a small portion around $\Omega + \omega_y$, effectively eliminating noise at $\Omega$ and $\Omega - \omega_y$, reduces the heating rate by a factor of 2 as compared to injecting the same amount of noise at $\Omega \pm \omega_y$.  This is consistent with $S_{V_\mathrm{RF}}(\Omega \pm \omega_y) = S_{V_\mathrm{RF}}(\Omega+\omega_y)+S_{V_\mathrm{RF}}(\Omega-\omega_y)$ for the case of broad spectrum noise \cite{Blakestad2011}.

To further demonstrate that the heating rates are unaffected by technical noise under the conditions of stray field compensation, heating rates were taken for several values of RF amplitude $V_0$.  Heating rates were found to be independent of voltage on the RF electrode, as shown in the inset of \figRef{fig:rfNoiseInjection}b.

Following Ref. \cite{Brownnutt2015}, the Johnson noise arising from the effective resistance of the RF resonator is estimated to be, $S_{V_\mathrm{RF}}(\Omega \pm \omega_y)  \sim 1 \times 10^{-17} \unit{V^2/Hz}$, orders of magnitude below the residual noise in our experiments ($> 1 \times 10^{-14} \unit{V^2/Hz}$) .  Note that the Johnson noise in this case is $\sim 5$ orders of magnitude larger than the DC case due to the large Th{\'e}venin equivalent resistance of the resonator. Near $\omega_y$ the Johnson noise on the RF electrode is estimated to be $S_{V_\mathrm{RF}}(\omega_y)  \sim 1 \times 10^{-20} \unit{V^2/Hz}$.  Due to the amount of Johnson noise and large characteristic distance of the RF electrode, $D_{y,\mathrm{RF}} > 1 \unit{m}$, there is not appreciable heating from noise on the RF electrode near $\omega_y$.   

\section{Conclusion}
We have presented a methodology for determining the contributions of technical noise, on both DC and RF electrodes, to the total motional heating rate in a surface-electrode Paul trap. The technique is based on injecting a known amount of noise onto the electrodes and measuring the corresponding increase in heating rate.  The contribution of the separately measured residual noise to the total heating rate can then be determined.  This method can be extended to other trap geometries and motional modes (including radial modes), providing a straightforward way to separate technical noise from other sources and aiding in the understanding of motional heating.   

Theoretical models are able to explain the increased heating rates from injected technical noise on the DC electrodes, and in particular the verification of the first term in \eqnRef{eq:omegaFits} allows calculation of expected heating.  These results may be instructive in the design of experiments that require fast shuttling of ions \cite{Walther2012} once the technical noise is directly measured. For example, suitable low-pass filtering of the DC voltages can be chosen such that the residual technical noise has a negligible contribution to the total heating rate, while fast shuttling times can still be achieved \cite{Kaufmann2014}.

Observation of increased axial heating rates in a linear trap due to excess noise on the RF electrode while the ion is displaced from the RF null illustrate the care that must be taken in the course of experiments.  The presence of a gradient large enough to cause significant heating in a mostly-symmetric trap highlights the need for careful minimization of the pseudopotential gradient to avoid excess motional heating. Additional filtering on the RF circuit, or the choice of a function generator with intrinsically lower noise is crucial, especially in cases where a pseudopotential gradient cannot be avoided.   

\section{Acknowledgments}
We thank Peter Murphy and Chris Thoummaraj for assistance with ion trap chip packaging and Vladimir Bolkhovsky for trap-chip fabrication. This work was sponsored by the Assistant Secretary of Defense for Research and Engineering under Air Force contract number FA8721-05-C-0002. Opinions, interpretations, conclusions, and recommendations are those of the authors and are not necessarily endorsed by the United States Government.

\bibliography{noisePaperBib}

\end{document}